\documentclass[
showpacs,
floatfix,
aps,
prl,
twocolumn,
superscriptaddress,
amssymb,
]{revtex4-1}

\usepackage{amssymb}
\usepackage{latexsym}
\usepackage[dvips]{graphicx}
\usepackage{amsmath}

\usepackage{graphicx}
\usepackage{dcolumn}
\usepackage{amsfonts}
\usepackage{bm}
\usepackage{epsfig}

\newcommand{\be}{\begin{equation}}
\newcommand{\ee}{\end{equation}}
\newcommand{\bea}{\begin{eqnarray}}
\newcommand{\eea}{\end{eqnarray}}

\newcommand{\pc}{{\mathcal P}}
\newcommand{\ac}{{\mathcal A}_0}

\def\Eac{\mathcal{E}}

\def\eac{\epsilon}

\def\epseff{\epsilon_{\mathrm{eff}}}

\def\oc{\omega_{\mbox{\scriptsize {c}}}}

\def\tq{\tau_{\rm q}}
\def\ttr{\tau}
\def\tem{\tau_{\rm em}}

\def\tee{\tau_{\it ee}}
\def\tin{\tau_{\rm in}}

\def\tst{\tau_\star}

\newcommand{\req}[1]{Eq.\,(\ref{#1})}

\newcommand{\rfig}[1]{Fig.\,\ref{#1}}

\newcommand{\rref}[1]{Ref.\,\onlinecite{#1}}

\begin{document}
\title{Microwave Photoresistance in an Ultrahigh Quality GaAs Quantum Well}
\author{Q.~Shi}
\affiliation{School of Physics and Astronomy, University of Minnesota, Minneapolis, Minnesota 55455, USA}
\author{S.\,A. Studenikin}
\affiliation{National Research Council of Canada, Ottawa, Ontario K1A 0R6, Canada}
\author{M.~A.~Zudov}
\email[Corresponding author: ]{zudov@physics.umn.edu}
\affiliation{School of Physics and Astronomy, University of Minnesota, Minneapolis, Minnesota 55455, USA}
\author{K.\,W. Baldwin}
\affiliation{Princeton University, Department of Electrical Engineering, Princeton, New Jersey 08544, USA}
\author{L.\,N. Pfeiffer}
\affiliation{Princeton University, Department of Electrical Engineering, Princeton, New Jersey 08544, USA}
\author{K.\,W. West}
\affiliation{Princeton University, Department of Electrical Engineering, Princeton, New Jersey 08544, USA}

\begin{abstract}
The temperature dependence of microwave-induced resistance oscillations (MIRO), according to the theory, originates from electron-electron scattering.
This scattering affects both the quantum lifetime, or the density of states, and the inelastic lifetime, which governs the relaxation of the nonequilibrium distribution function.
Here, we report on MIRO in an ultrahigh mobility ($\mu > 3 \times 10^7$ cm$^2$/Vs) 2D electron gas at $T$ between $0.3$ K and $1.8$ K. 
In contrast to theoretical predictions, the quantum lifetime is found to be $T$-independent in the whole temperature range studied.
At the same time, the $T$-dependence of the inelastic lifetime is much \emph{stronger} than can be expected from electron-electron interactions.

\end{abstract}
\maketitle

Microwave-induced resistance oscillations (MIRO) appear in two-dimensional electron \citep{zudov:2001a,ye:2001} or hole systems \citep{zudov:2014,shi:2014b} subjected to low temperature $T$, weak magnetic field $B$, and radiation of frequency $f =\omega/2\pi$. 
When the MIRO amplitude exceeds dark resistance, the oscillation minima evolve into zero-resistance states \citep{mani:2002,zudov:2003,yang:2003,willett:2004,smet:2005,zudov:2006a,zudov:2006b,bykov:2006,konstantinov:2010,hatke:2012b}, understood in terms of formation of current domains \citep{andreev:2003,auerbach:2005,finkler:2009,dmitriev:2013,dorozhkin:2011,dorozhkin:2015}.

When the microwave power is not too high and Landau levels are overlapping, away from the cyclotron resonance MIRO can be described by \citep{dmitriev:2012}
\be
\delta R  \approx - A\sin 2\pi\eac\,,~A = \eac p(\eac) \lambda^2 \ac \,,  
\label{eq.miro}
\ee
where $\eac=\omega/\oc$, $\oc=eB/m^\star$ is the cyclotron frequency, $m^\star$ is the effective mass, $p(\eac)$ is a dimensionless function describing absorption \citep{note:p}, $\lambda=\exp(-\pi/\oc\tq)$ is the Dingle factor, $\tq$ is the quantum lifetime,  and 
\be
\ac = 2 \pi R_0 \bar{\pc} \left(\tau/2\tst+2\tin/\tau \right)\,.
\label{eq.miro2}
\ee
Here, $R_0$ is the resistance at $B=0$, $\bar{\pc}$ is the dimensionless microwave power \citep{note:pc}, $\tau$ is the transport lifetime, $\tst$ is a disorder-specific scattering time \citep{note:7}, and $\tin$ is the inelastic lifetime.
The first term in \req{eq.miro2} describes the  \emph{displacement} contribution \citep{ryzhii:1970,ryzhii:1986,durst:2003,lei:2003,vavilov:2004,dmitriev:2009b}, owing to the radiation-induced modification of impurity scattering. 
The second term represents the  \emph{inelastic} contribution \citep{dorozhkin:2003,dmitriev:2003,dmitriev:2005}, originating from the radiation-induced oscillations in the electron distribution function.

It is well established that the MIRO amplitude $A$ decreases with $T$ \citep{zudov:2001a,mani:2002,zudov:2003,studenikin:2005,studenikin:2007,hatke:2009a,wiedmann:2010a}.
This decrease can originate from $\ac$, through $\tin(T)$, and/or from $\lambda$, through $\tq(T)$. 
One study \citep{hatke:2009a} has found that, at $T \gtrsim 1$ K, $\lambda$ decreased considerably, while $\ac$ remained essentially unchanged \citep{note:2}.
Such a behavior can indeed be expected at high $T$, when the $T$-independent displacement contribution [first term in \req{eq.miro2}] dominates over the inelastic contribution [second term in \req{eq.miro2}], which is predicted to decay as $T^{-2}$.
However, at low $T$, when $\tq$ becomes $T$-independent, the $T$-dependence of $A$ should be dominated by $\ac(T)$.
In the limit of low $T$, one thus expects  $A \sim \ac \sim T^{-2}$, as the inelastic contribution should overwhelm the $T$-independent displacement contribution.

While the above prediction has been known for more than a decade, it remained unclear if it could ever be tested experimentally.
The lack of such experiments, at least in part, is due to cHallenges associated with combining low temperature and microwave radiation of intensity sufficient for reliable MIRO detection.
According to \req{eq.miro}, the MIRO amplitude depends exponentially on the quantum lifetime entering the Dingle factor. 
This fact dictates that exploration of the low temperature limit calls for samples of exceptional quality.

In this Rapid Communication we report on microwave photoresistance measurements in an ultra-clean two-dimensional electron gas with mobility $\mu > 3 \times 10^7$ cm$^2$/Vs and quantum mobility $\mu_{\rm q}\equiv e\tq/m^\star > 1 \times 10^6$ cm$^2$/Vs at temperatures between $T \approx 0.3$ K and $T \approx 1.8$ K.
At low microwave intensities, we observe surprisingly strong temperature dependence of the MIRO amplitude.
In qualitative agreement with the above prediction, we find that the dependence originates from the $T$-dependent $\ac$, while $\lambda$, parametrized by $\tq \approx 46$ ps \citep{note:1}, remains constant.
However, the $T$-dependence of $\ac$ is significantly \emph{stronger} than theoretically predicted.
At the same time, estimates show that $\tq$ should still be $T$-dependent, in contrast to our observations.
These findings might indicate that electron-electron scattering becomes ineffective at low temperatures.
Phenomenologically, the $T$-dependence of the MIRO amplitude can be well described by a simple exponential function.

Our sample is a $4\times4$ mm square cleaved from a symmetrically doped, 30-nm wide GaAs/AlGaAs quantum well. 
After low-temperature illumination with a red light-emitting diode, electron density and mobility were $n_e \approx 3.2 \times 10^{11}$ cm$^{-2}$ and $\mu \approx3.1 \times 10^7$ cm$^2$/Vs, respectively.
Microwave radiation of frequency $f = 34$ GHz \cite{note:111} was delivered to the sample via a semirigid coaxial cable terminated with a 3 mm antenna \citep{bogan:2012}.
The longitudinal resistance $R$ was measured using a low-frequency (a few Hz) lock-in amplification under continuous microwave irradiation, in sweeping magnetic field, and at temperatures, measured by a calibrated thermometer, $T$ between $0.3$ and $1.8$ K.

\begin{figure}[t]
\includegraphics{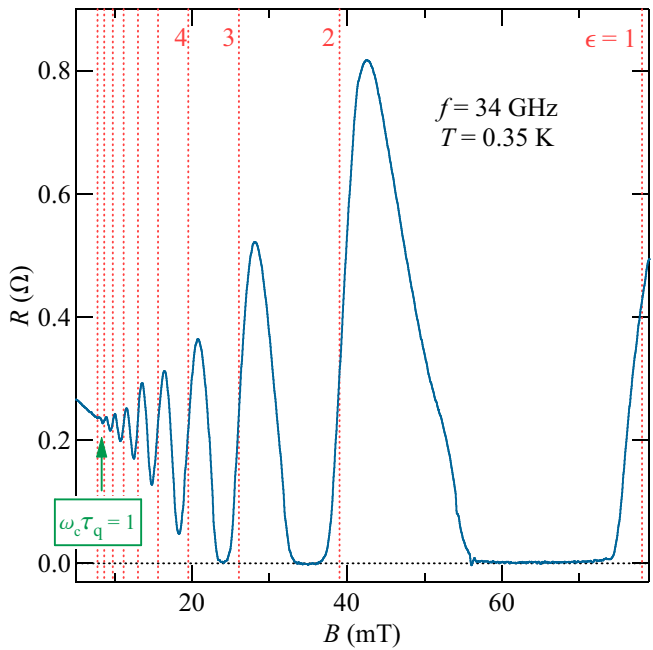}
\vspace{-0.15 in}
\caption{(Color online)
Magnetoresistance $R(B)$ under irradiation by microwaves at $T = 0.35$ K. 
Vertical lines are drawn at $\eac = \omega/\oc = 1,2,..., 10$.
Arrow is drawn at $B = 8.3$ mT, marking the condition $\oc\tq = 1$.
}
\label{fig1}
\end{figure}

In \rfig{fig1} we present magnetoresistance $R(B)$ measured under irradiation by microwaves at $T = 0.35$ K. 
Vertical lines are drawn at $\eac = \omega/\oc = 1,2,..., 10$.
The data reveal multiple MIRO and zero-resistance states near $\eac = 1,2$ and 3.
We ascribe the exceptional quality of our data to the extremely high value of the quantum lifetime, $\tq \approx 46$ ps.
This value translates to quantum mobility of $\mu_{\rm q} \approx 1.2 \times 10^6$ cm$^2$/Vs which, to our knowledge, is the highest value reported to date in any system.
The arrow marks $\oc\tq = 1$ and roughly corresponds to the onset of MIRO.

\begin{figure}[t]
\includegraphics{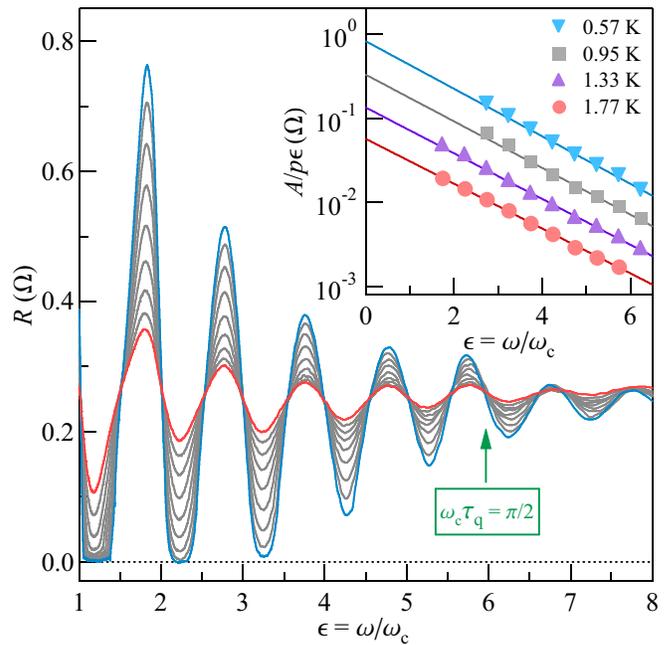}
\vspace{-0.15 in}
\caption{(Color online)
$R$ vs $\eac$ at temperatures from $T = 0.48$ K (largest amplitude) to $T = 1.25$ K (smallest amplitude).
Arrow is drawn at $\eac = 6$, marking the condition $\oc\tq = \pi/2$.
Inset shows $A/p\eac$ at different $T$ (see legend) vs $\eac$ on a semi-log scale. 
Solid lines are fits to $\ac \exp(-\eac/f\tq)$.
}
\label{fig2}
\end{figure}

In \rfig{fig2} we show $R$ vs $\eac$ at $T = 0.48$, 0.57, 0.66, and from 0.75 to $1.25$ K, in a step of 0.1 K.
We immediately observe that the $T$-dependence is not only significant, but also uniform over the whole range of $\eac$.
The continuous change of the MIRO amplitude indicates that the electron temperature does not fully saturate due to heating by microwave radiation down to $T \approx 0.5$ K.
We note that, according to the theory \citep{ando:1974b}, the gap between Landau levels opens when $\oc\tq = \pi/2$, which corresponds to $\eac = 6$, marked by $\uparrow$.
No qualitative difference in the data obtained in the regimes of overlapping and separated Landau levels is observed \cite{hatke:2011f,hatke:2012D}.

\begin{figure}[t]
\includegraphics{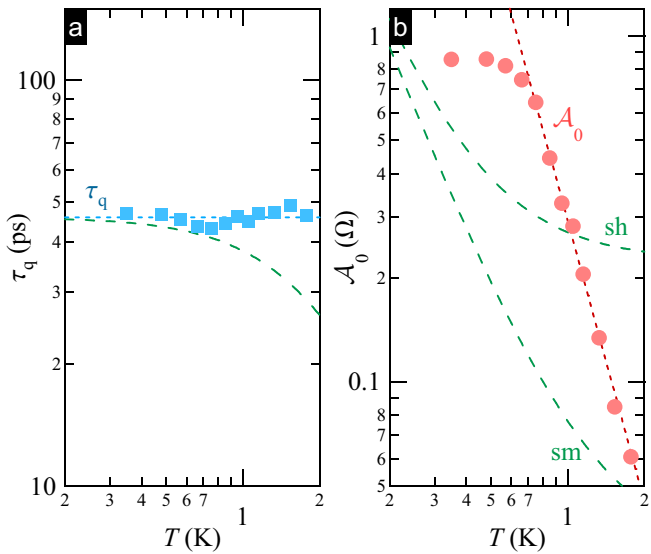}
\vspace{-0.15 in}
\caption{(Color online)
(a) $\tq(T)$ (squares).
Dotted line represents the average $\tq = 45.9$ ps.
Dashed line illustrates theoretical dependence, see text.
(b) $\ac(T)$ (circles).
Dotted line is $\sim T^{-2.7}$.
Dashed lines are calculated $T$-dependencies of $\ac$ for sharp (upper) and smooth (lower) disorder limits \citep{note:7}.
}
\label{fig3}
\end{figure}

We start the data reduction by constructing Dingle plots which allow us to assess the $T$-dependencies of the Dingle factor $\lambda$ and of
\be
\ac = \lim_{\eac \to 0} A/p\eac\,.
\ee
In the inset of \rfig{fig2} we show Dingle plots for several temperatures, see legend.
We find that the data are well described by exponential dependencies with approximately the same slope, indicating that $\tq$ has very weak $T$-dependence. 
On the contrary, $\ac$, given by the intercept,  changes significantly with $T$. 
Indeed, as illustrated in \rfig{fig3}(a), $\tq$ (squares) is roughly $T$-independent and is close to $\tq \approx 46$ ps.
At the same time, \rfig{fig3}(b) reveals that $\ac$ (circles) exhibits very strong $T$-dependence changing by more than an order of magnitude. 
This change occurs despite the apparent saturation at lower $T$ (to which we will return later).
We thus conclude that the MIRO temperature dependence originates primarily from $\ac(T)$ and not from $\lambda(T)$.
We note that this conclusion is opposite to that of \rref{hatke:2009a} which examined MIRO in a lower quality sample ($\mu \approx 1.3 \times 10^7$ cm$^2$/Vs, $\mu_{\rm q} \approx 0.5 \times 10^6$ cm$^2$/Vs) and at higher $T$, from 1 to 4 K.

We next compare the $T$-dependence of $\ac$ to the theoretical predictions. 
The first term in \req{eq.miro2}, the displacement contribution, is $T$-independent and its value is governed by the correlation properties of the disorder potential.
For purely smooth disorder, $\ttr/2\tst = 6/(\ttr/\tq+3)$ \citep{dmitriev:2005,dmitriev:2009b}, which is about 0.21  in our sample.
In the sharp disorder limit, $\ttr/2\tst$ increases to $1.5$ \citep{dmitriev:2009b}.
The second term in \req{eq.miro2} represents the inelastic contribution and is $T$-dependent.
When $\pi T \gg \omega$, \rref{dmitriev:2005} predicts 
\be
\tin \approx 0.82 \tee\,,~~ \frac \hbar \tee = \frac{\pi k_B^2 T^2}{4 E_F}\ln\left( \frac{2 \hbar v_F}{a_B k_B T}\right)\,,
\label{eq.tee}
\ee
where $\tee^{-1}$ is the electron-electron scattering rate for a test particle at the Fermi energy $E_F$, $v_F$ is the Fermi velocity, and $a_B \approx 10$ nm is the Bohr radius in GaAs.
At $T=1$ K, the logarithmic factor is about 6 and we estimate $\tin \approx 0.18$ ns, leading to $2\tin/\ttr \approx 0.3$ in our sample.
This value is comparable to the displacement contribution in the smooth disorder limit.

Based on the above estimates, we expect $\ac$ to increase by a factor 
between 1.3 (sharp disorder limit) and 2.3 (smooth disorder limit) as the $T$ is lowered from 1.77 to 0.75 K.
However, the data in \rfig{fig3}(b), reveal a considerably larger, more than an order of magnitude increase of $\ac$.
The discrepancy between experiment and theory can be evaluated via a direct comparison of the measured $\ac(T)$ and calculated $T$-dependencies. 
The latter are shown by dashed lines in \rfig{fig3}(b) for the limits of sharp (upper curve) and smooth (lower curve) disorder.
Since any real sample has both sharp (unintentional background impurities) and smooth (remote ionized donors) disorder, one should expect a $T$-dependence lying between the two curves.
However, even if one assumes the limit of smooth disorder, the experimental $T$-dependence is markedly stronger than theoretically predicted.
In particular, at $T > 0.7$ K, one finds an empirical relation, $\ac \sim T^{-2.7}$, shown by dotted line.
\begin{figure}[t]
\includegraphics{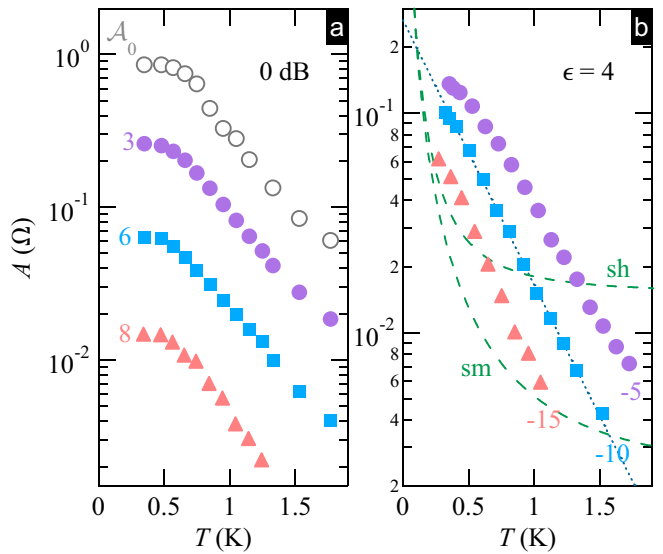}
\vspace{-0.15 in}
\caption{(Color online)
(a) $A(T)$ at $\eac = 3, 6, 8$ (solid symbols, as marked) and $\ac(T)$ (open symbols) at 0 dB attenuation.
(b) $A(T)$ at $\eac = 4$ and at a power attenuated by $-5$ (circles), $-10$ (squares), and $-15$ dB (triangles).
Dashed lines are calculated $T$-dependencies of $\ac$ for sharp (upper) and smooth (lower) disorder limits \citep{note:9}.
Solid line is $\sim \exp(-T/T_0)$, where $T_0 = 0.35$ K.
}
\label{fig4}
\end{figure}

We now turn our attention to the saturation of $\ac(T)$ observed at low $T$ in \rfig{fig3}(b). 
In \rfig{fig4}(a) we plot $\ac(T)$ (open symbols) together with the MIRO amplitude $A$ (solid symbols) at $\eac = 3, 6, 8$ on a semi-log scale.
We observe that all $T$-dependencies are roughly parallel to each other, which is indeed expected for a $T$-independent $\lambda$.
In addition, we find that all dependencies slow down at $T\lesssim 0.7$ K.We note that \rref{dmitriev:2005} predicts a crossover from a $T^{-2}$-dependence of the inelastic relaxation rate to a $T^{-1}$-dependence as the $T$ is lowered below $\hbar\omega/\pi k_B \approx 0.5$ K.
However, a slowdown of the $T$-dependence can also result from an experimental difficulty to cool electrons under the presence of radiation.

To test the second scenario we have performed additional measurements (in a separate cooldown), utilizing lower microwave intensities.
The results of these measurements are presented in \rfig{fig4}(b) showing the $T$-dependencies of the MIRO amplitude at $\eac = 4$ for three different powers.
These powers are lower than the power used earlier by $-5$ (circles), $-10$ (squares), and $-15$ dB (triangles).
Direct comparison with the 0 dB data shown in \rfig{fig4}(a) reveals that the tendency to saturate at low $T$ is greatly reduced at lower powers.
We thus conclude that unintentional heating of electrons by microwave radiation is the main cause of the slowdown observed in \rfig{fig3}(b) and \rfig{fig4}(a) at low $T$.

At lower microwave intensities, the disagreement between experiment and theory becomes more pronounced.
The calculated ratio of the MIRO amplitudes at 0.5 K and 1.5 K is between 1.6 (sharp disorder, upper curve) and 3.7 (smooth disorder, lower curve).
Both these values are much smaller than the ratio of more than 15 observed in the $-10$ dB data.
We note that even at these much lower powers the electron temperature can still be higher than the thermometer reading and that accounting for this error would only increase the discrepancy.
We would also like to comment that the MIRO $T$-dependence appears to be well described by an exponential function, as illustrated in \rfig{fig4}(b) by a dotted line $\sim \exp(-T/T_0)$, with $T_0 =  0.35$ K.

We now return to $\tq(T)$ shown in \rfig{fig3}(a).
At finite $T$, the theory \citep{dmitriev:2009b,dmitriev:2012} dictates that the density of states acquires an interaction-induced correction and that $\tq$, entering $\lambda^2$, is given by
\be
1/\tq = 1/\tq^0+\lambda'/\tee\,,
\label{eq.tq}
\ee
where $\tq^0 = \tq(T=0)$, $\lambda'\sim 1$ \citep{dmitriev:2009b}, and $\tee$ is given by \req{eq.tee}.
With $\tq^0 = 46$ ps, $\lambda' = 1$, we calculate $\tq(T)$ and present the result in \rfig{fig3}(a) as a dashed curve.
In contrast to the measured $\tq(T)\approx {\rm const}$ (squares), calculated $\tq(T)$ decreases by $\sim 40$\% in the experimental $T$-range.
It would indeed be interesting to extend experiments to higher $T$ to see whether or not $\tq$ eventually decreases.
Unfortunately, such studies are not feasible for two reasons.
First, even at the highest power [see \rfig{fig4}(a)], MIRO decay very rapidly with $T$ \citep{note:10}, which progressively limits the accuracy of the Dingle analysis.
Second, in ultra-high mobility samples, such as ours, phonon-induced resistance oscillations \citep{zudov:2001b,zhang:2008,hatke:2011d} set in at $T <$ 2 K \citep{hatke:2011d}.
These oscillations grow with $T$, interfering with and further obscuring the MIRO signal.
As a result, we are not able to draw any conclusions about the fate of $\tq$ at higher $T$.
However, the disagreement with the theory is already evident from the data acquired at accessible temperatures.

It is interesting to examine possible reasons causing apparent disagreement between the present study and that of \rref{hatke:2009a}.
Since the mobility of our sample is considerably higher than in \rref{hatke:2009a}, and since the mobility is believed to be limited by sharp disorder \citep{umansky:1997,umansky:2009,manfra:2014}, the displacement contribution should be much smaller in our experiment.
This conclusion is consistent with our observations that (a) MIRO cease to exist at much lower $T$ than they do in \rref{hatke:2009a} and (b) $\ac$ shows much stronger $T$ dependence compared to \rref{hatke:2009a}.
The disagreement in $\tq(T)$ likely originates from the fact that \rref{hatke:2009a} has focused on higher $T$ and, as a result, has not examined the $T$ range below 2 K in detail \citep{note:11}. 
However, later experiments on nonlinear transport in the same sample \citep{hatke:2009c,note:12} and another study \citep{dietrich:2012} found saturation of $\tq$ at low $T$, in agreement with our findings. 

Another question is why the observed $T$-dependence of $\ac$ ($\tq$) is much stronger (weaker) than theoretically predicted.
Phenomenologically, our findings might indicate that, for some reason, electron-electron scattering is much less effective than expected.
If so, one would naturally expect a stronger $\ac(T)$ in our ultra-high mobility sample, where the displacement contribution is suppressed, and
a weaker $\tq(T)$.
We note that such a scenario has been recently proposed to explain faster than $T^{-2}$-dependence of the inelastic relaxation at low $T$ \citep{dietrich:2015,note:13}.

In summary, we have studied the temperature dependence of MIRO in an ultrahigh quality GaAs quantum well.
We have found that, in contrast to theoretical predictions, the quantum lifetime remains essentially unchanged.
Nevertheless, the temperature dependence of the MIRO amplitude is significant and originates primarily from a much faster than expected $T$-dependence of the inelastic contribution.
At the same time, the displacement contribution appears to be unusually weak in our sample.
Taken together, these findings appear to cHallenge our current understanding of microwave photoresistance and call for further investigations.

\begin{acknowledgments}
We thank I. Dmitriev and D. Polyakov for discussions. 
The work at Minnesota was funded by the NSF Grant No. DMR-1309578.
Q.S. acknowledges Allen M. Goldman fellowship.
The work at Princeton was partially funded by the Gordon and Betty Moore Foundation and by the NSF MRSEC Program through the Princeton Center for Complex Materials (DMR-0819860).
Preliminary measurements were performed at the National High Magnetic Field Laboratory, which is supported by NSF Cooperative Agreement No. DMR-0654118, by the State of Florida, and by the DOE.
\end{acknowledgments}




\begin{thebibliography}{62}
\bibitem[{not({\natexlab{c}})}]{note:7}
\bibinfo{note}{Lines are drawn at $a(6/[\ttr/\tq+3]+2\tin/\ttr)$ (lower) and
  $a(3/2+2\tin/\ttr)$ (upper), where $a = 0.15$.}
\bibitem[{not({\natexlab{e}})}]{note:1}
\bibinfo{note}{This value was obtained from the low-temperature Dingle analysis
  of MIRO measured at $f = 34$ GHz.}
\bibitem[{not({\natexlab{g}})}]{note:9}
\bibinfo{note}{Lines are drawn at $a(6/[\ttr/\tq+3]+2\tin/\ttr)$ (lower) and
  $a(3/2+2\tin/\ttr)$ (upper), where $a = 0.01$.}
\bibitem[{not({\natexlab{h}})}]{note:10}
\bibinfo{note}{Already at $T \approx 1.8$ K, $A \sim 0.01~\Omega$ at $\eac = 3$
  and is even smaller at higher $\eac$.}
\end{thebibliography}

\end{document}